# Detection Method Based on Automatic Visual Shape Clustering for Pin-Missing Defect in Transmission Lines

Zhenbing Zhao, *Member IEEE*, Hongyu Qi, Yincheng Qi, Ke Zhang, *Member IEEE*, Yongjie Zhai, *Member IEEE*, Wenqing Zhao

*Abstract*—Bolts are the most numerous fasteners in transmission lines and are prone to losing their split pins. How to realize the automatic pin-missing defect detection for bolts in transmission lines so as to achieve timely and efficient trouble shooting is a difficult problem and the long-term research target of power systems. In this paper, an automatic detection model called Automatic Visual Shape Clustering Network (AVSCNet) for pin-missing defect is constructed. Firstly, an unsupervised clustering method for the visual shapes of bolts is proposed and applied to construct a defect detection model which can learn the difference of visual shape. Next, three deep convolutional neural network optimization methods are used in the model: the feature enhancement, feature fusion and region feature extraction. The defect detection results are obtained by applying the regression calculation and classification to the regional features. In this paper, the object detection model of different networks is used to test the dataset of pin-missing defect constructed by the aerial images of transmission lines from multiple locations, and it is evaluated by various indicators and is fully verified. The results show that our method can achieve considerably satisfactory detection effect.

*Index Terms*—Transmission line, bolt, pin-missing defect, visual shape, deep convolutional neural network.

## I. INTRODUCTION

IN transmission lines, bolts as a kind of fasteners are widely applied to the connection between various components of transmission lines, making the whole structure stable. However, due to the complex working environment, they are prone to breakage. Some pins may be missing, which may cause a large area of transmission line fault and seriously threatens the safety and stability of the grid. For traditional bolt inspections, climbing up to the positions is the main means to check the bolts, which is time-consuming and laborious. Due to the scattered distribution of bolts and the variety of bolt specifications, the bolt inspection becomes more difficult [1], [2]. In recent years, the Unmanned Aerial Vehicle transmission line inspection has been promoted in the power system for their high security and efficiency, which also can combine with object detection technology of machine learning to realize intelligent processing [3]-[6]. Fig. 1 is the schematic diagram of the aerial photography process of the transmission line, and the main shooting areas are Area 1, Area 2 and Area 3. The shooting instance is shown in Fig. 2.

This work is supported in part by the National Natural Science Foundation of China under Grant 61871182, Grant 61401154, Grant 61773160, and Grant 61302163, by Beijing Natural Science Foundation under Grant 4192055, by the Natural Science Foundation of Hebei Province of China under Grant F2016502101, Grant F2017502016, and Grant F2015502062, by the Fundamental Research Funds for the Central Universities under Grant 2018MS095, Grant 2018MS094, and by the Open Project Program of the National Laboratory of Pattern Recognition under Grant 201900051. (Corresponding author: Zhenbing Zhao).

Z. Zhao, H. Qi, Y. Qi, K. Zhang are with the School of Electrical and Electronic Engineering, North China Electric Power University, Baoding 071003, China (e-mail: zhaozhenbing@ncepu.edu.cn; nansbas@163.com; qiych@126.com; zhangke41616@126.com).
Y. Zhai are with the department of Automation, North China Electric Power University, Baoding 071003, China (e-mail: zhaiyongjie@ncepu.edu.cn).
W. Zhao are with the School of Computer Science and Technology, North China Electric Power University, Baoding 071003, China (e-mail: zhaowenqing@ncepu.edu.cn).

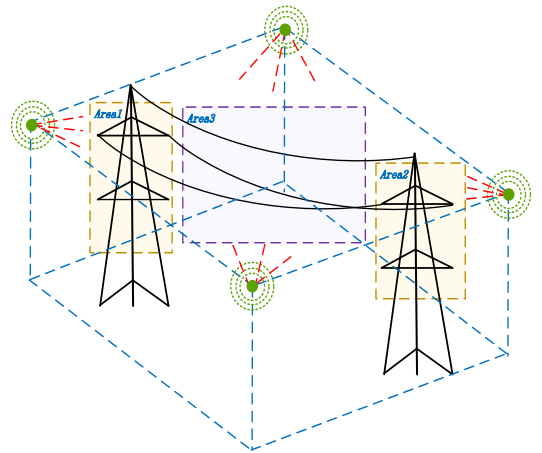

Fig. 1. Acquisition of aerial images of transmission lines.

Due to the complex environments and various structures of transmission lines, the vision-based bolt pin-missing detection is influenced by some interference caused by unpredictable complex background changes and the shielding by other components of the line. The bolt is physically small compared with other components in transmission lines, making it extremely difficult to obtain single bolt aerial image by air vehicles and leading to the extreme imbalance between object information and non-object information at the same time. It further increases the difficulty in detecting the pin-missing defect. Even worse, the visual difference between the pin-missing bolt and the normal one in aerial images is extremely little, and the irregularity of the defect features makes it more challenging to classify the defect features of pin-missing bolts in aerial images. Traditional object detection methods based on artificial feature design confront great limitations when dealing with the problems above [7]. When the three-dimensional information of the bolt is converted into the two-dimensional images, different two-dimensional bolt visual structures will be generated, which makes the visual representation of pin-missing bolt inconsistent and interferes with the learning process of visual features.

In this paper, to solve the inconsistent interference of the two-dimensional visual structure of bolt, an automatic visual shape clustering method is proposed and a two-stage object detection model is developed based on Faster R-CNN. In order to effectively extract the small-scale bolt features in aerial images, three key improvements are implemented on developed model, including the feature enhancement, feature fusion, and expansion regions of interest (RoI) feature extraction. These improvements are not independent but obtained by the comprehensive design considering the prior characteristics of the defects mentioned in this paper. Finally, a model for pin-missing detection of bolts in transmission lines is constructed which called AVSCNet.

The main contributions of this paper are as follows:

1) An automatic visual shape clustering method is proposed for bolts in transmission lines. Based on the method, an object detection model based on deep convolutional neural network which is called AVSCNet, is constructed to adapt to the characteristics of pin-missing defect detection in the aerial images of transmission lines.

2) Aiming at the ambiguous correspondence relation between the high-level semantic features of the network and the pixel area of the visual word in the original image, an expansion RoI feature extraction method is proposed and applied to AVSCNet. The method can optimize the radiation range of the feature information and reduce the quantization error to improve the effectiveness of features of small pin-missing objects.

3) A feature enhancement method based on bilinear interpolation and a feature fusion method are applied to AVSCNet. These two optimizations work together to combat the inertia of excessive information corrosion in deep networks and improve the ability of feature discernment in small pin-missing defect detection.

Four mainstream models and multiple indicators are applied to provide objective comparison and evaluations for the research.

The rest of the paper is organized as follows. Section II analyzes the related work of this paper and summarizes the main difficult points encountered during the research. In Section III, the methods proposed in this paper and the key steps of our proposed model in detail are illustrated; meanwhile, the principle analysis of the multiple optimizations is also made. In Section IV, we introduce comparison models, data sets, and experimental configuration. In Section V, the experimental results are given and comparative analysis is made. Finally, in Section VI, some conclusions are drawn and summary of this paper is given, and in addition, the direction and suggestions are put forward for future work.

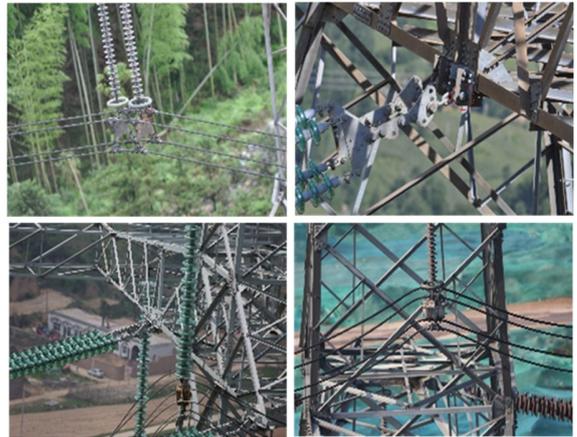

Fig. 2. Aerial images of complex transmission lines.

## II. Research Background

Object detection is one of the most basic and challenging problems in computer vision. Its task is to locate the object of predefined categories from natural images [8]. In recent years, because of the wide application of deep learning, the object detection field has developed rapidly [9]-[11]. Different from traditional methods such as SIFT, SURF and BRISK [12]-[15], the deep convolutional neural network (DCNN) has the ability of extracting abstract features of data essentially, which makes it widely studied and applied in the field. The main object detection framework based on DCNN can be divided into two categories in terms of their structures: the two-stage detection framework based on region proposal, and the one-stage detection framework based on regression. Faster R-CNN [16] is the typical one for the former, while SSD [17] and YOLO [18]-[20] are the representatives of the latter.

In recent years, there have been many studies on the application of automatic defect detection based on computer vision. In [21], an automatic detection system for fastener defects on railway rails was proposed. The line segment detector was used to locate the rail sleepers, and the fasteners were structurally modeled. Finally, the Haar features of the fastener objects were used for fault recognition. In [22], computer vision was adopted to detect the icing phenomenon of cables on the transmission lines. The author proposed the Canny edge detection algorithm based on Support Vector Regerssion (SVR), and then combined it with the Hough and grayscale statistical features to realize icing anomalies detection.

Based on the DCNN, Reference [23] proposed an unsupervised texture surface defect detection method, and constructed a multi-scale convolutional neural network to extract features and inversely reconstruct surface defects, which was more complicated. Small pixel-by-pixel visual defect detection in the complex background was realized. For the contact network on high-speed railway lines, a three-stage defect detection structure based on SSD and YOLO was built in [24]. Firstly, the fasteners were detected after the support device was located. After that, the defects were detected by constructing a deep neural network. The entire automatic detection of fastener defects on the catenary supporting devices can be realized. Compared with [24], Reference [25] further studied the pin-missing defects on the catenary supporting devices. The author constructed an optimized PVANET structure to unify the original neural network structure of different network structures in [24], and changed the last stage from identification to the detection of multiple local regions. The proposed method combined the visual information of several regions to judge the defect of the pin and can obtain good effect. In [26], the detection of up to 12 components on the high-speed rail contact network was carried out, and the performance of multiple deep models in this scenario was compared, and various component defects were detected further.

The object detection based on deep learning is also applied to many other practical scenarios. A detection method was proposed in [27] for surface aging of railway track. It obtained good detection results for the minor aging defects by improving the regularization algorithm of the deep model and using data enhancement. In [28], the convolutional neural networks were applied to detect one-class defect of product, and a loss function based on Euclidean distance penalty was designed for defect recognition, which achieves satisfactory defect detection results and has strong generalization. In [29], the Faster R-CNN was applied to detect the scratches and points on the wheel hub in an image with complicated background. By using the convolutional neural network to detect pattern defects and identify minor defects at the same time, the higher defect detection accuracy was achieved. Aiming at a variety of typical road defects, a multi-scene road defect detection deep model was constructed through the convolutional neural network in [30].

The object detection method based on computer vision is widely applied in many fields and solves many different problems, but the research on the bolt defect detection of the transmission line has been rarely studied by far. There are still many questions to be solved in the application of deep learning-based computer vision methods in the pin-missing detection of transmission lines.

1) There are huge differences in the visual shapes of multi-angle shooting bolts, which may lead to poor similarity of visual words.

2) The backgrounds in the aerial images are complex and diverse, thus higher requirements for notable features to distinguish from other objects are demanded.

3) Bolts are comparatively small in the transmission lines and have less absolute pixel information. When the deep structure extracts image features, the abstract semantic features of the object will be extracted, and the information of the object will also be corroded.

4) The data-driven deep learning method needs a lot of supervised data to support the sensitive perception ability of the model, but it is difficult to acquire the supervised aerial images. This job requires not only the abundant experiences of professional inspection staff, but also the semantic rationality from the perspective of computer vision. There is a semantic gap between the two terms.

To realize the comprehensive and accurate positioning of bolts in aerial images and the accurate classification of the bolts, the following work should be done. Firstly, a dataset with high-quality supervision information is the prerequisite and should be constructed to support the research. Secondly, in the case of finite data, it is necessary to construct the features with strong robustness and category discrimination ability. The method proposed in this paper combines the characteristics of the bolt and to a certain extent eliminates the negative impact of non-object factors in the complex environment of the transmission line.

III. AVSCNET STRUCTURE AND PRINCIPLE

The basic framework of AVSCNet is depicted in Fig. 3, and the data flow process of training and testing is given separately. For better illustration, the abstract network parameters are used to describe the network structure of the training, and a reified feature map is utilized to show the working process during the testing. The figure highlights our proposed automatic visual shape clustering process and three optimization implementations: the feature enhancement, feature fusion and expansion region feature extraction. The clustering part of the training framework is unsupervised, while the part after clustering is supervised learning; the two parts are an undividable whole. The principle based on the training framework of Fig. 3 will be elaborated below.

*A. Automatic Visual Shape Clustering Method*

An important part of supervised learning is to define the cognitive data to be used in the model. The artificial data supervision is based on the comprehensive cognition constructed based on the long-term accumulation of multi-modal information of the object, regardless of the model characteristics. The most direct manifestation is the high degree of abstraction of the visual description of one-type objects. The visual shapes of different objects at different angles are defined as a same visual word. The difference of the visual shapes of bolts is particularly noticeable in the aerial images of the transmission lines. As shown in Fig. 4, when the bolt in the three dimensions is represented by a two-dimensional image, it presents a variety of visual shape differences due to the angles of the photographing and incomplete symmetry structure of the bolt itself.

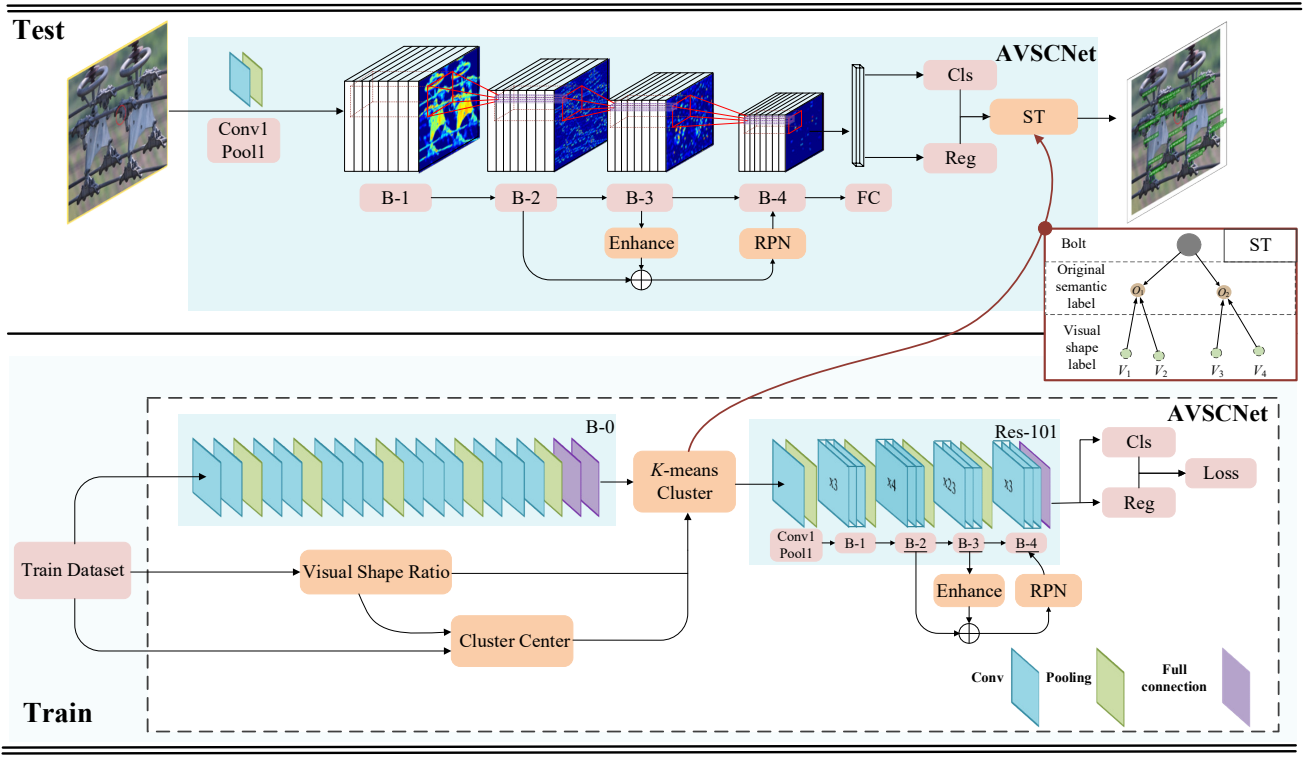

Fig. 3. Overview of AVSCNet.

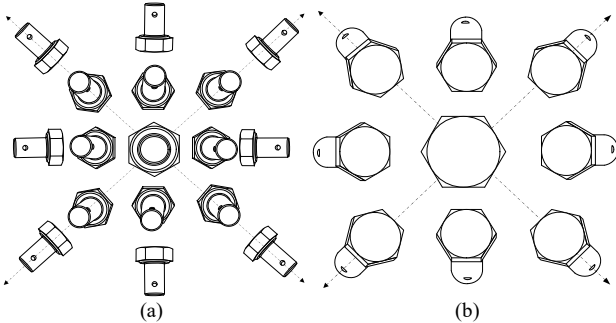

(a)                         (b)
Fig. 4. Schematic diagram of the difference in visual shape of bolts in different angles. (a) a series of bolt shape facing the tail of the bolt, and (b) a series of bolt shape facing the head of the bolt.

The multiple-visual-shape discrimination of bolts is not a difficult task for humans, but for the deep model, things are quite opposite and the difference in visual contents between the same types becomes a heavy burden. The contradiction of data formed by too many intra-class data difference will mislead the learning process of the neural networks. In addition to the differences in visual contents, the statistical distribution in the height and width of objects in different visual shapes also demonstrate their own characteristics.

The height and width of the $i$-th sample $x_i$ of one class in the defined dataset $\chi$ are $H_i$ and $W_i$, and the visual shape ratio of a single object is $\xi_i$. Each object of the class in dataset can be calculated by equation (1):

$$\xi_i = \frac{\min[H_i, W_i]}{\max[H_i, W_i]} \quad (1)$$

The deep model pre-trained in the ImageNet large-scale dataset is used as the general feature extractor, and $\psi(\cdot)$ is defined as the deep network feature extraction. For the input object image $x_i$, the extracted visual shape optimization features are calculated by equation (2):

$$f_i = \xi_i \times \psi(x_i) \quad (2)$$

Where $\psi(x_i)$ represents the deep feature extraction by the pre-trained DCNN, and $f_i$ represents the optimized deep feature obtained by the joint optimization of visual shape ratio and pre-trained DCNN.

As shown in Fig. 5(a), the curve of visual shape ratio distribution of all bolt samples is visualized after kernel density estimation by the Gaussian kernel function. There are two obvious peaks in the visual shape parameters of bolts. Considering the distance between the peaks, the width of density interval is defined as 0.2, and for each extreme interval value $d_v$, the morphological quantity $Mu$ can be calculated by equation (3).

$$Mu = \sum_{v=1}^{t} \rho[d_{v-1} < d_v \mid d_{v+1} < d_v] \quad (3)$$

Where $\rho[\cdot]$ is the identity function with the value of 1. The interval from 0 to the maximum of the shape ratio of all the data is evenly divided into $t$ parts. In this paper, $t = 5$. By equation (3), all subintervals with the most concentrated samples within the interval are found, and then the total number of these subintervals is taken as the number of clustering centers. Thus the Object visual shape clustering is automatically implemented by clustering the optimized visual shape deep features using $K$-means. At the same time, the relation tree between the visual-shape semantics and the original semantics is called ST which contains the connection relations between the original labels and the corresponding multiple visual-shape semantic labels. When the image is input, the model will directly output a visual shape label of the detected object, and then convert the visual shape label into the original semantic label by ST. As shown in Fig. 3, in the ST, $O_1$ represents pin missing semantic, and $V_1$ and $V_2$ respectively represent the classification of two visual shape labels of the pin-missing bolt. $W_i$ and $H_i$ of different shape categories are plotted as the two-dimensional curves for visual observation. The clustering results are shown in Fig. 5(b). It can be observed that the automatic visual shape clustering results are basically consistent with those in Fig. 5(a).

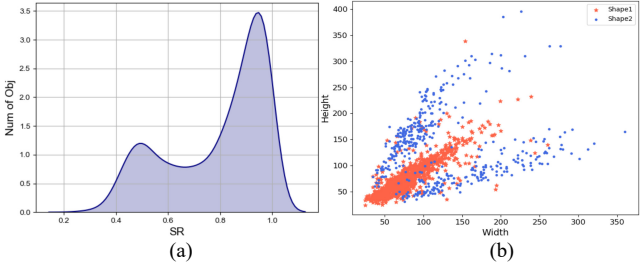

Fig. 5. (a)Visual shape ratio kernel density estimation based on Gaussian kernel. (b) Objects length and width distribution of different visual shape.

### B. Feature Extraction and Feature Enhancement

(1) Aerial image feature extraction

In this paper, the hybrid structure based on Vgg-16 [31] and ResNet-101 [32] is used as the backbone. The overall structure parameters of the network are listed in TABLE I. Each part of the model is described in detail as follows: "size" indicates the size and number of the convolution kernel; "quantity" indicates the number of convolution operations of the corresponding size.

TABLE I
PARAMETERS OF AVSCNET NETWORK

| Items | B-0 | B-1 | B-2 | B-3 | B-4 |
|---|---|---|---|---|---|
| Size | 3×3×(64, 128,256, 512,512) | 1×1/64 3×3/64 1×1/256 | 1×1/128 3×3/128 1×1/512 | 1×1/256 3×3/256 1×1/1024 | 1×1/512 3×3/512 1×1/2048 |
| Quantity | 2 | 3 | 4 | 23 | 3 |

Among them, B-1~B-4 are the residual blocks.

After the visual shape clustering, the size of each image in the training dataset is reshaped to 900/720 pixels to ensure the stable computation process. Then the image is divided into multiple batches with the size of 128. In the process of each batch flowing into B-1~B-3 and flowing out of B-3, the convolution, pooling and ReLU activation are operated and finally the deep features of the aerial images are obtained.

(2) Feature enhancement

The flow of data in the deep network shows a trend of channel increasing and feature map size decreasing. One of the reasons is that the pooling layer exists in convolutional neural network. The process of feature extraction can be explained as a process of corroding the original image with purpose and directivity, by which the key features can be effectively extracted to get the correct classification and regression results. But the network itself cannot automatically adapt to the process, for it has been set in the design process. For relatively small objects, the deeper networks may yield worse results. In order to extract the high-level semantic features, the features are excessively corroded. As a result, the amount of information related to the object is too less, which fundamentally affects the effectiveness of the obtained features and also makes a decrease in the discriminability, and eventually leads to the occurrence of false and missed detection.

Let the size of the feature map input to the $i$-th convolution layer be $w_{i-1} \times h_{i-1}$, the size of the convolution kernel of this layer be $k_{i-1}$, the convolution stride of horizontal and vertical be $s_1$ and $s_2$ respectively, and the length of the edge-filled pixels be $p$. The width $w_i$ and height $h_i$ of the output feature map of $k_{i-1}$ layer output are calculated as follows:

$$w_i = (w_{i-1} + 2*p - k)/s_1 + 1 \qquad (4)$$

$$h_i = (h_{i-1} + 2*p - k)/s_2 + 1 \qquad (5)$$

As can be seen from the above equations, when the step size is greater than 1, there are two trends as the network deepens: the decreasing resolution and the less activated area. The visualized results of the image of the data flowing through the model are demonstrated in Fig. 6. The aerial image of the ResNet-101 network is demonstrated in Fig. 6(a), while Fig. 6(b), Fig. 6(c), and Fig. 6(d) are the results of the up-sampling of the first three block outputs of ResNet-101. When seeking the correspondence between the object semantics and the object visual content region, the ideal object semantic response region and the object visual content have an exact one-to-one correspondence. In practice, the relationship is rather ambiguous. The visual information of many bolt objects had been exceedingly corroded by deep networks and submerged in layers deep into the model. The loss of the correspondence makes it impossible to detect the object.

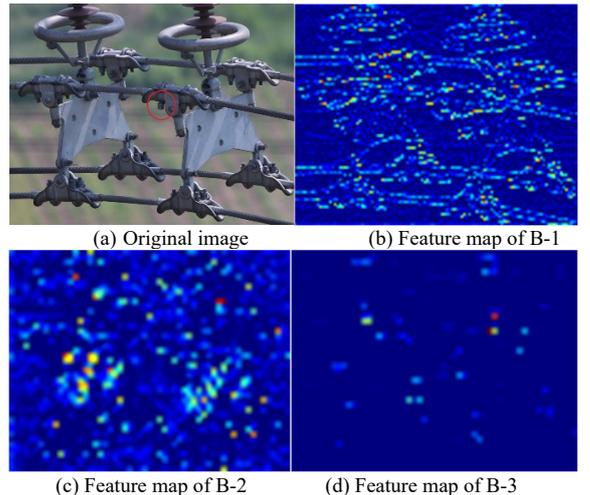

Fig. 6. Feature sparse process of extracting semantic features from deep networks.

In order to improve the model to adapt to the small object detection, the bilinear interpolation is used to enhance the resolution of features in this paper. Let $F(w_i, h_j)$ be the feature value of the points of $(w_i, h_j)$ positions on the feature map, and

the interpolated values of any position $F(w_i+u, h_j+v)$ of floating point type are calculated as follows:

$$F(w_i + u, h_j + v) = \frac{F(w_i, h_j) + F(w_{i+1}, h_j) + F(w_i, h_{j+1}) + F(w_{i+1}, h_{j+1})}{4} \quad (6)$$

Through the feature enhancement, the features become clearer and more prominent. In the next section, we will introduce multi-scale feature fusion. Combined with the method in this section, the deeper network can also produce the deep features with clearer correspondence to the original object, and the highly abstract semantic features can be obtained.

*C. Region Proposal Based on Feature Fusion*

The excellent feature extraction ability of deep learning benefits from the continuous accumulation of multi-layer distributed feature expression. Different feature extraction layers in the network construct a hierarchically rich visual pattern extraction structure, and the rich network nodes in the same layer fully express the specific visual pattern. A large number of research results show that the features from the deep layer of the network have abstract semantics, and the feature layer at the front of the network has more clear correspondence with the original data pixel information, while the semantic features at the high level show a relatively ambiguous correspondence. The high-level semantic features are obtained by convolution and nonlinear activation of low-level detail features, but it does not mean that high-level semantic features are the sum of low-level features and cannot contain all the important information of low-level features. To sum up, there is a certain contradiction between the abstract semantic features and the low-level features. The former may lack enough information to support their meanings due to the over-abstract simplification, while the latter ones contain a great amount of redundant information.

Whether the small objects are absolutely or relatively small in the image should be considered as a unified problem in the model, which is too little pixel information in the object area can be found. The high-level semantic feature cannot describe the object completely, while the redundant information of the low-level feature will cause excessive interference in the process of the model perceiving the object. This problem is particularly obvious in the inspection task of bolt defect in the aerial images. Through the feature enhancement method, the overall feature resolution of aerial images is effectively enhanced to ensure the full capacity of feature space, but the feature cannot be enriched. In order to resolve the contradiction between the two and obtain more discriminating features for small objects, this paper adopts a multi-scale feature fusion method to construct a deep model. Combining the B-2 output features with the enhanced B-3 output features through the concatenate method, it not only retains the abstraction of high-level semantic features, but also suppresses the noise redundancy of low-level features to some extent. The features of the original ResNet output in B-3 have 1024 convolution kernels, and a total of 1536 convolution kernels are used to express the features more abundantly through the feature fusion layer.

The anchor mechanism is applied and the anchor ratios are set to 0.5, 1, 2, and 3 with multiple scales 64, 128, 256, and 512. For each anchor, 16 anchor boxes of different sizes are used to calculate the confidence degree of foreground and background in the fusion feature. The bounding box regression is subsequently implemented, and the region proposals can be obtained eventually.

*D. Classification and Regression Based on Expansion RoI*

(1) Expansion RoI

RoI pooling is a general operation for extracting features of RoIs of Regional Proposal Network (RPN) [17]. It first rounds the coordinates of the continuous floating-point foreground region calculated based on the RPN to obtain the position information, and then the features of corresponding positions are extracted from the input features. Through the maximum pool processing, the regional features of different sizes will be converted to the fixed sizes, which is the basis for the detection of objects with different scales. Anchor boxes corresponding to objects of different scales can be classified and regressed in this part. The integer adjustment of the coordinate with floating type will make the feature region inconsistent with the actual one. The minor errors introduced in the quantization process have little impact on the case that the content is clearly visible and the structural characteristics are obvious, but they will affect the subsequent classification and regression of the objects in small areas. For the problem of quantization error, RoI Align method was proposed in [33], which fundamentally changes the coordinate position processing of the selected RoI sampling region. For small objects, the number of regional features mapped to the last layer is often too small, and the information is dispersed into nearby feature values during the process of convolution and pooling instead of being fully utilized.

In view of this, we propose the expansion RoI (ERoI) layer to eliminate the problem of small object feature loss caused by the quantization of RoI Pooling, and fully extract the deep features of the foreground. As shown in Fig. 7, the differences of the three methods are exhibited by calculating the average value of the foreground features. In Fig. 7, for the areas containing more object information, the features are effectively extracted by using ERoI via bilinear interpolation; in such way, the deep visual features of small objects are well retained. Our method takes the characteristics of receptive field overlap of convolutional neural network into account; the most relevant spatial contextual information is added to the small object feature calculation process to provide more sufficient and effective information for the classification and regression calculation.

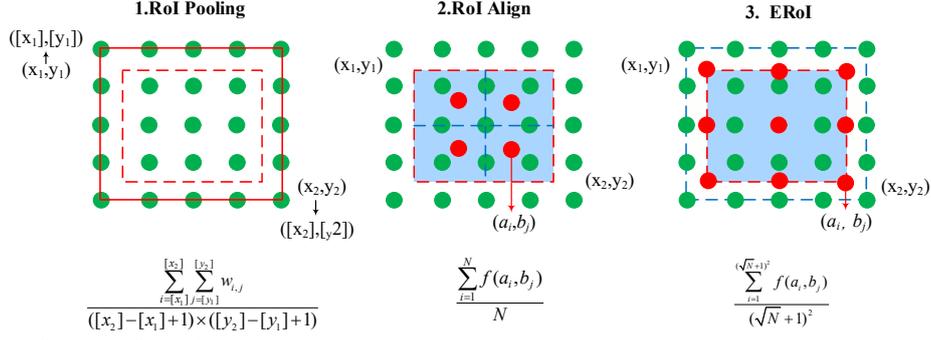

Fig. 7. Illustration of RoI Pooling, RoI Align and ERoI.

For different sizes of RoIs, a 7×7 fixed-size RoI is used for feature extraction. Firstly, the feature area of 14×14 is obtained by ERoI, and then the maximum pooling is performed with the stride step of 2. Finally, the area feature with the size of 7×7×1536 is obtained. The error loss of the learning process is calculated by equation (7), wherein the RPN loss is calculated in the same manner as the classification regression process after B-4. The loss after B-3 is taken as an example, and the calculation equation is as follows.

$$L(\{p_i\},\{t_i\}) = \frac{1}{N_{cls}} \sum_i L_{cls}(p_i, p_i^*) + \lambda \frac{1}{N_{reg}} \sum_i p_i^* L_{reg}(t_i, t_i^*) \quad (7)$$

Here, $i$ is the suggested index and $p_i$ is the predicted probability of the index $i$ being an object of one class; $p_i^*$ is the ground truth; $t_i$ is a vector representing four parameterized coordinates of the predicted bounding box and $t_i^*$ is the real box associated with $t_i$. Both $L_{cls}$ and $L_{reg}$ functions used in equation (7) follow the description in [16], and the two terms are normalized by $N_{cls}$ and $N_{reg}$, and a balancing weight $\lambda$ is added. $L(\{p_i\}, \{t_i\})$ is the final loss obtained after classification and regression.

(2) Classification and Regression

A 7×7×1536 tensor is extracted by ERoI and it flows through a fully connected layer containing 1000 nodes. For $K$ classes of objects to be detected, the feature vectors with the length of 1000 are respectively connected to the fully connected layer of length ($k+1$) and the fully connected layer of 4×($k+1$). They correspond to the number of categories containing the background and the position correction offset to be calculated, respectively. After the Softmax function calculation and position regression calculation, the classification probability and object position correction are obtained. The non-maximum suppression (NMS) is adopted for multiple redundant boxes of objects, which limits the redundant bounding boxes by considering the overlapping areas and scoring of different bounding boxes. Some examples of the final test results are shown in Fig. 8, where the red rectangular box represents the defect of pin-missing bolts, the red circle represents the defect area confirmed by the power inspection professionals, and the green rectangular box represents the normal bolt with pins. After the redundant object proposal is removed by NMS, different visual patterns of different categories will pass through ST, enabling the model to return the detection results consistent with the semantics of the root node.

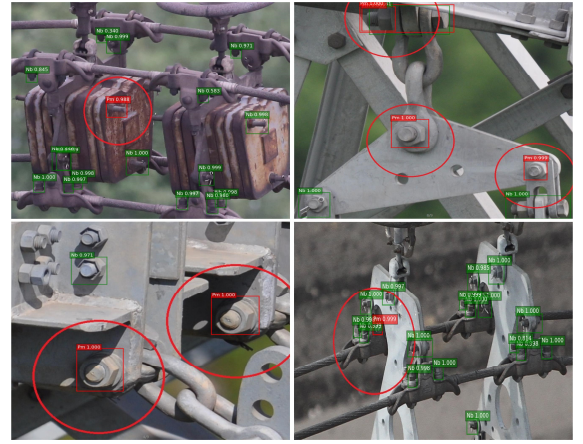

Fig. 8. Detection results of AVSCNet based on test set.

In the training process, the classification loss and regression loss after the model RPN and B-4 are calculated in the same way as B-3, and the cross-entropy cost function and the Smooth L1 function [8] are applied respectively. The total loss of the training network is the sum of classification loss and regression loss calculated in RPN and after B-4 feature extraction.

IV. DATA SET AND EXPERIMENTAL SETUP

A. Dataset and Server Configuration

1) Dataset: There are 1843 aerial images in our defect detection system, including a total of 6880 bolt objects. The data were captured by air vehicles from different power lines in multiple locations. In order to improve the generalization ability of the learning model, we randomly scrambled the aerial images of different angles in different sections of different places. The ratio of training data to test data is set to 8:2, and 1471 images are used to train while 369 images are used to test. The number of objects used during training and testing are listed in TABLE II.

TABLE II
TRAINING AND TESTING DATASET

| Bolt States | Images Num | Normal bolt(Nb) | Pin missing(Pm) |
|---|---|---|---|
| Train | 1471 | 4111 | 1292 |
| Test | 369 | 1073 | 404 |
| Total | 1840 | 5184 | 1696 |

The LabelImg software is used to mark all the bolts in the aerial images except for the bolts on the tower, for the bolts on the tower have no pin missing defect, which is determined when it was designed. Generally, the bolts can be divided into two categories, one is normal and the other is pin-missing. The former comprises a nut, a cotter pin and a screw; the latter lacks the split pin, which becomes its main characteristic. After the data labeling work is completed, the bolts in the images have only one single visual word. This dataset is used to complete the training and testing of the model for the detection of pin-missing defects.

2) Server configuration: The experimental GPU server environment for the model proposed in this paper is configured as follows: the software frame is TensorFlow, and the operation system is Linux Ubuntu16.04; its CPU is E5-2620 v4 at 2.10 GHz, and other hardware configurations are a 32-GB RAM, and NVIDIA TITAN Xp GPU with 12-GB memory.

B. Experimental Setup

In this part, the most representative YOLOv3 in the single-stage object detection framework and the most representative Faster R-CNN and R-FCN models in the two-stage object detection framework are selected for comparison. In order to fully verify the proposed method, different depths of ResNet structure are used to support our comparison experiment. Before the comparison, the following brief introduction is given for better illustration.

1) ResNet-50: The overall structure of ResNet-50 is the same with ResNet-101 in TABLE I. Except the number of residual blocks corresponding to B-3 which is reduced from 23 to 6; that is, the convolution layer is 51 layers less, lighter than ResNet, which results in a faster detection response.

2) YOLOv3: The multiple 1×1, 3×3 convolution kernels are the backbone [17] of the model DarkNet-53. YOLOv3 implements multi-scale prediction on different layer feature layers. By intensive sampling uniformly and using different scales and aspect ratios in different positions of the image and then using CNN to extract features and implement direct classification and regression, the model is proved to be simple and efficient. The area to be processed contains a large body of backgrounds, which makes the positive and negative samples extremely unbalanced during the training. In fact, it is the root cause of the difficulty in improving the accuracy of such frameworks.

3) R-FCN is an optimized detection model based on Faster. In the RoI feature processing part after RPN, the sensitive ROI-Pooling method and the position-sensitive scoring graph voting mechanism are proposed. The convolution is more sensitive to the location of the object; besides, the use of the global average strategy to replace the fully connected layer after RoI pooling can speed up the process.

4) Feature Pyramid Network (FPN) [34] is a kind of convolutional neural network structure. Faster-FPN is an FPN-based Faster R-CNN which uses a pyramidal multi-scale approach in the feature construction and feature application. For the definition of multi-scale here has two meanings, it is multi-scale of the feature structure which constructs fusion features of different resolutions; the other is that the multi-scale prediction is applied. In this paper, we use the Faster-FPN based on ResNet-50 and ResNet-101 for comparison.

C. Training Process

The momentum optimization method is adopted in the training process, and the step learning rate is set. The learning rate is attenuated at 6w, 9w, and 11w, and the initial learning rate is 0.001; the attenuation coefficient is 0.1, and a total of 13w iterations are performed. The momentum is set to 0.9, the IoU threshold $a_{min}$ equals 0.5, and the batch size is set to 128 after RPN and B-4. The NMS threshold applied to RPN proposals is set to 0.7. The convergence performance of the model of the entire training process of AVSCNet is expressed by the overall loss, which can be calculated by equation (7). As shown in Fig. 9, the abscissa is drawn with 20 times compression for easy viewing, and the red triangle indicates the specific location that the learning rate starts decreasing.

D. Evaluation Indexes

In the contrast experiments, the average precision rate (AP) and average recall rate (AR) were used to evaluate the performance of the model, and the mean AP (mAP) and mean AR (mAR) are used to evaluate the overall performance of the model. The data of the test set were input onto the trained model to perform forward propagation, and the area of the predicted bounding box is $S_{pre}$ and the area of the ground true box is $S_{gt}$. If the condition of the following equation is satisfied, it can be determined as a detected object. The process can be illustrated by equation (8).

$$\frac{S_{pre,ij} \cap S_{gt,ij}}{S_{pre,ij} \cup S_{gt,ij}} > a_{min} \qquad (8)$$

Where $a_{min}$ is the minimum intersection over union (IoU) threshold. After comparing the classification structure of the object with the label of Ground True, the detection results are divided into four categories: the true positive (*TP*), false positive (*FP*), true negative (*TN*), and false negative (*FN*). *Precission* (*P*) and *Recall* (*R*) are calculated as follows.

$$P = \frac{TP}{TP+FP} \times 100\% \qquad (9)$$

$$R = \frac{TP}{TP+FN} \times 100\% \qquad (10)$$

The mAR can be used to evaluate the overall performance of the model for object classification, it is the mean of AR. AR calculated as follows.

$$AR = \frac{\sum_{i=1}^{Q_{cls}} R_i}{Q_{cls}} \qquad (11)$$

Where $Q_{cls}$ represents the total number of detected objects of one class.

The mAP can be used to evaluate the overall performance of the classification and regression of the test model, which is calculated as follows.

$$\text{mAP} = \frac{\sum_{i=1}^{N_{cls}} \int_0^1 P_i(R_i)dR}{N_{cls}} \quad (12)$$

## V. COMPARISON OF TESTING RESULTS

The parameters of anchor mechanism are the same with the training ones. Since bolts are relatively small objects in aerial images, the model is prone to missing detecting them, so $a_{min}$ is set to 0.5. The P-R curves for all models are divided into two parts, one is based on the smaller backbone, including ResNet-50 and DarkNet-53; the other is a deep network structure based on ResNet-101, as shown in Fig. 10 and Fig. 11 respectively. The mAP can be used to evaluate the performance of model classification and regression as a whole, and it can be calculated by equation (12). By using AR, a better comparison of positioning performance of the model can be made, calculated by equations (10) and (11). The results of various experiments are analyzed, shown as follows.

1) Loss: As can be seen from the loss curve of AVSCNet based on ResNet-101 in Fig. 9, the whole process of convergence is accompanied by drastic changes. The difference between normal and pin-missing is subtle. Besides, the factors such as the shooting angles, complex background, occlusion, shadow, blurring and others also put great pressure on the classifier, thus exacerbating the loss.

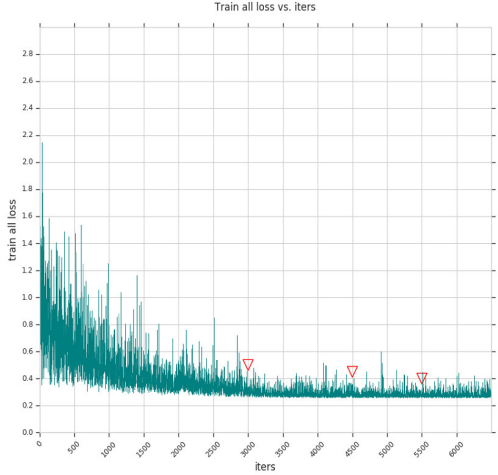

Fig. 9. Loss changes during training.

2) In order to verify the effectiveness of the various network optimization methods of this paper more intensively, the ablation experiment was made on the AVSCNet with three different optimization method. Since the feature fusion method based on up-sampling was adopted in this paper, feature enhancement (FE) is required if feature fusion (FF) is adopted. The results of experiment are listed in TABLE III and TABLE IV. The left side of the table represents the methods used. The experimental results show the impact of different strategies on the performance of AVSCNet, among which FF contributes significantly to the improvement of AVSCNet performance.

TABLE III
THE CONTRIBUTION OF EACH ELEMENT IN OUR DETECTION PIPELINE. THE BASELINE MODEL IS RESNET-50 AVSCNET

| FF | FE | ERoI | AP | | mAP | AR | | mAR |
|---|---|---|---|---|---|---|---|---|
| | | | Nb | Pm | | Nb | Pm | |
| | | | 0.676 | 0.645 | 0.661 | 0.854 | 0.815 | 0.835 |
| | ✓ | | 0.685 | 0.656 | 0.671 | 0.862 | 0.824 | 0.843 |
| ✓ | ✓ | | 0.702 | 0.665 | 0.684 | 0.875 | 0.838 | 0.857 |
| | | ✓ | 0.684 | 0.647 | 0.666 | 0.860 | 0.816 | 0.838 |
| ✓ | ✓ | ✓ | **0.706** | **0.667** | **0.687** | **0.878** | **0.845** | **0.862** |

TABLE IV
THE CONTRIBUTION OF EACH ELEMENT IN OUR DETECTION PIPELINE. THE BASELINE MODEL IS RESNET-101 AVSCNET

| FF | FE | ERoI | AP | | mAP | AR | | mAR |
|---|---|---|---|---|---|---|---|---|
| | | | Nb | Pm | | Nb | Pm | |
| | | | 0.692 | 0.677 | 0.685 | 0.863 | 0.842 | 0.853 |
| | ✓ | | 0.705 | 0.685 | 0.695 | 0.869 | 0.852 | 0.861 |
| ✓ | ✓ | | 0.725 | 0.710 | 0.718 | 0.881 | 0.869 | 0.875 |
| | | ✓ | 0.698 | 0.678 | 0.688 | 0.867 | 0.847 | 0.857 |
| ✓ | ✓ | ✓ | **0.734** | **0.714** | **0.724** | **0.889** | **0.862** | **0.876** |

3) In the detection of the normal bolts and the ones with pin-missing defect, the best AR and AP are obtained by our model. A variety of feature optimization methods used in the construction of the model proposed in this paper are implemented on AVSCNet, allowing for more detailed comparison. The speed of AVSCNet is slower than others, and the time consumed by our model is 6 times that of YOLOv3. It can be found from TABLE V that the extra time is mainly spent on the feature enhancement and feature fusion. The former sacrifices efficiency in order to make the correspondence between the feature area and the original object area clearer; while the latter increases the number of feature channels in order to strengthen the detailed information and restrain the excessive semantic feature.

TABLE V
MODEL VERIFICATION RESULTS BASED ON RESNET-50 AND DARKNET-53

| Methods | AP | | mAP | AR | | mAR | FPS |
|---|---|---|---|---|---|---|---|
| | Nb | Pm | | Nb | Pm | | |
| YOLOv3 | 0.536 | 0.559 | 0.548 | 0.678 | 0.656 | 0.667 | 11.36 |
| Faster R-CNN | 0.639 | 0.588 | 0.614 | 0.806 | 0.717 | 0.762 | 5.80 |
| R-FCN | 0.644 | 0.634 | 0.639 | 0.818 | 0.743 | 0.781 | 8.99 |
| Faster-FPN | 0.689 | 0.655 | 0.672 | 0.864 | 0.833 | 0.849 | 3.05 |
| AVSCNet | **0.706** | **0.667** | **0.687** | **0.878** | **0.845** | **0.863** | 1.92 |

4) The comparison of all models on mAP is shown in TABLE VI. The AVSCNet based on ResNet-101 achieved the highest pin-missing detection result of 71.4%. The accuracy of the normal category is generally higher than the defect category, for the normal bolts in transmission lines are always the majority, while the defect samples are relatively little, resulting in the sample imbalance during the training. It is natural for the model that produces a preference for the category which accounts for the largest proportion of all the samples.

TABLE VI
MODEL VERIFICATION RESULTS BASED ON RESNET-101

| Methods | AP | | mAP | AR | | mAR | FPS |
| --- | --- | --- | --- | --- | --- | --- | --- |
| | Nb | Pm | | Nb | Pm | | |
| Faster R-CNN | 0.658 | 0.603 | 0.631 | 0.827 | 0.754 | 0.791 | 3.09 |
| R-FCN | 0.665 | 0.645 | 0.641 | 0.839 | 0.788 | 0.814 | 5.26 |
| Faster-FPN | 0.716 | 0.692 | 0.704 | 0.872 | 0.851 | 0.862 | 2.21 |
| AVSCNet | **0.734** | **0.714** | **0.724** | **0.889** | **0.862** | **0.876** | 1.02 |

5) Fig. 10 demonstrates the performance of different models. The performance of the multiple two-stage model constructed with a 50-layer convolution residual network is significantly better than that of YOLOv3. This is because the detection idea adopted by YOLOv3 pursues efficiency, and the classification and regression processing for the generated default box are not filtered. The positioning effect with large image resolution and small object area is relatively poor, and the small category difference increases the burden.

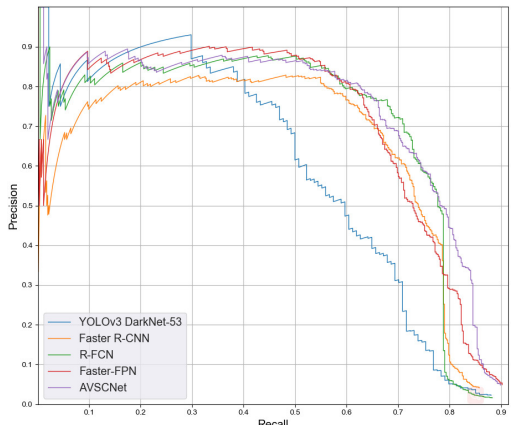

Fig. 10. P-R curve of Pm based ResNet-50 and DarkNet-53.

6) A comparison and visualization of the RP curves of all pin-missing detection is demonstrated in Fig. 11 based on Resnet-101 backbone model. ResNet-101 trades deeper networks for more discriminating features, so most of the curves are more complete and better than Fig. 10. The precision of the method proposed in this paper appears a rapid decline near the place that *Recall*=0.8. Because some of the features of small objects have been corrupted thoroughly by the network, they are not visible in the final classification and regression. The large receptive field will aggravate the problem, essentially limit the possibility of detection, and show that there are difficult samples in the dataset, which is quite natural.

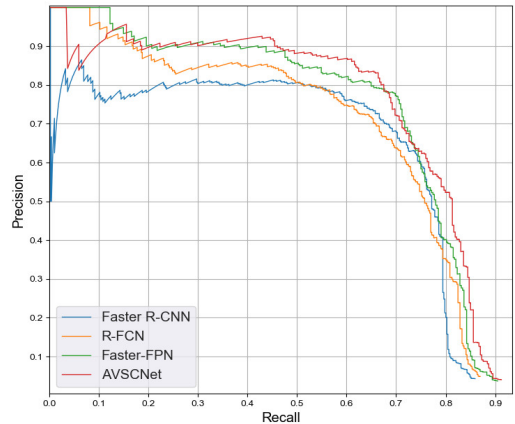

Fig. 11. P-R curve of Pm based ResNet-101.

## VI. CONCLUSION

In order to detect the pin-missing defect of bolts in transmission lines with good accuracy, a new visual shape clustering method for bolts is presented in this paper. A detection model called AVSCNet is developed based on automatic visual shape clustering method and three optimization methods of DCNN. Three kinds of evaluation indicators are used for performance evaluation of the multiple-object detection model based on deep learning. The experimental results show that the model can effectively detect the pin-missing defect of bolts in transmission lines, and eliminate the non-uniform interference generated when the three-dimensional bolt entity is mapped to the two dimensional images; besides, the corrosiveness of deep network to the small bolt objects is reduced, and thus the small bolt feature expression is optimized. However, there are still some problems to be solved:

1) Distributional difference: Even if the differences in the shooting equipment are not considered, the diversity of the transmission line structures and natural environments in different locations makes the aerial image data a dependent-identity distribution, which leads to insufficient guarantee of the availability of the model to more extensive transmission lines under the finite training set and a strong dependence on the supervised data. It should be pointed out that the acquisition of the supervised data is not easy. More and more images of the transmission lines will be taken, which makes the problem more and more prominent.

2) Model limitations: the deep model can perceive the image information well by constructing the effective correlation between the feature expression and the supervision information. However, the bolt defect object is different from the general visual ones which not only depends on the visual content of the object area, but also depends on the structural information of the scene where the object is located. For example, a normal bolt with a pin should be distinguished from a normal bolt without a pin through different local scenarios in a transmission line.

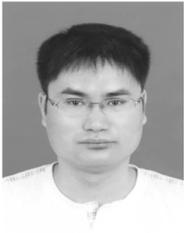 **Zhenbing Zhao** was born in Jiangsu, China in 1979. He received his B.S. degree in 2002, M.S. degree in 2005 and PhD in 2009 from North China Electric Power University, Baoding, Hebei. Now he is an associate professor in School of Electrical and Electronic Engineering, North China Electric Power University. His research interests include artificial intelligence, intelligent detection of electrical equipment and image processing.

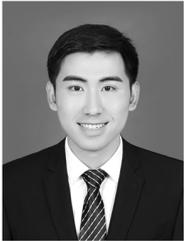 **Hongyu Qi** was born in 1994. He received the bachelor degree from North China Electric Power University, Baoding, Hebei, China in 2017. He is currently a second year postgraduate with department of Electronics major, School of Electrical and Electronic Engineering, North China Electric Power University, Baoding, Hebei, China. His research interests include machine learning and digital image processing.

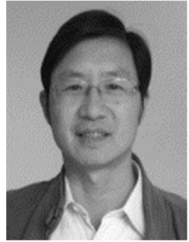 **Yincheng Qi** was born in 1968. He received his B.S. degree in 1990, M.S. degree in 1998 and PhD in 2009 from North China Electric Power University, Baoding, Hebei. Now he is a professor in North China Electric Power University. His research interests include electric power system communication and information processing.

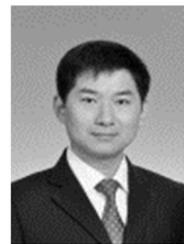 **Ke Zhang** was born in 1980. He received his B.S. degree in 2003, M.S. degree in 2006 from North China Electric Power University, Baoding, Hebei. And received his PhD degree from Beijing University of Posts and Telecommunications, Beijing. Now he is an associate professor in School of Electrical and Electronic Engineering, North China Electric Power University. His research interests include deep learning, robot navigation, natural language processing and spatial relation description.

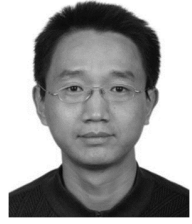 **Yongjie Zhai** received the Ph.D. degree from North China Electric Power University, China in 2004. He is currently a professor with department of Automation, North China Electric Power University, China. His research interests include pattern recognition, digital image processing and computer control system.

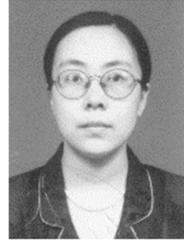 **Wenqing Zhao** was born in Shanxi province, China. She received Master degree of computer application in 1999 from North China Electric Power University, and received Doctor degree of Electric Engineering and Automation in 2009 from North China Electric Power University. At present she is a professor in North China Electric Power University. Her present research interests are artificial intelligence and deep learning.